\documentstyle[twoside,fleqn,espcrc2]{article}

\input{psfig}

\newcommand{\Z}{{\sf Z \!\!\! Z}}

\newcommand{\Psibar}{\bar{\Psi}}

\title{Flop Transitions in Cuprate and Color Superconductors: \\
From $SO(5)$ to $SO(10)$ Unification?}

\author{S. Chandrasekharan\address{Department of Physics, Duke University, 
Durham, NC 27708,U.S.A.}, V. Chudnovsky\address{Center for Theoretical Physics,
Laboratory for Nuclear Science and Department of Physics \\ Massachusetts 
Institute of Technology (MIT), Cambridge, MA 02139, U.S.A.}, 
B. Schlittgen$^{{\rm b}}$ and U.-J. Wiese$^{{\rm b}}$}

\begin{document}

\begin{abstract}

The phase diagrams of cuprate superconductors and of QCD at non-zero baryon
chemical potential are qualitatively similar. The N\'eel phase of the cuprates 
corresponds to the chirally broken phase of QCD, and the high-temperature 
superconducting phase corresponds to the color superconducting phase. In the 
$SO(5)$ theory for the cuprates the $SO(3)_s$ spin rotational symmetry and the 
$U(1)_{em}$ gauge symmetry of electromagnetism are dynamically unified. This 
suggests that the $SU(2)_L \otimes SU(2)_R \otimes U(1)_B$ chiral symmetry of 
QCD and the $SU(3)_c$ color gauge symmetry may get unified to $SO(10)$. 
Dynamical enhancement of symmetry from $SO(2)_s \otimes \Z(2)$ to $SO(3)_s$ is 
known to occur in anisotropic antiferromagnets. In these systems the staggered 
magnetization flops from an easy 3-axis into the 12-plane at a critical value 
of the external magnetic field. Similarly, the phase transitions in the $SO(5)$
and $SO(10)$ models are flop transitions of a ``superspin''. Despite this fact,
a renormalization group flow analysis in $4-\epsilon$ dimensions indicates that
a point with full $SO(5)$ or $SO(10)$ symmetry exists neither in the cuprates 
nor in QCD.

\end{abstract}

\maketitle

\newpage

Understanding QCD at non-zero baryon chemical potential $\mu$ is very important
in the context of both heavy ion and neutron star physics. While asymptotically
large values of $\mu$ are accessible in perturbative QCD calculations, such 
values are not realized in actual physical systems. Investigations of the 
phenomenologically relevant regime at intermediate $\mu$ require the use of 
nonperturbative methods. Unfortunately, first-principles lattice calculations 
in this regime are presently prevented by the notorious complex action problem.
Conjectures for the QCD phase diagram at non-zero $\mu$ are thus based on model
calculations. These calculations reveal interesting phenomena such as color
superconductivity \cite{Bai84} but one cannot expect the results to be 
quantitatively correct. 

While it is very important to develop quantitative methods to understand QCD at
non-zero $\mu$, here we ask if further qualitative insight can be gained 
through analogies with related condensed matter systems. In particular, the 
phase diagram of high-temperature cuprate superconductors is qualitatively 
similar to the one conjectured for two flavor QCD. The ordinary hadronic phase 
of QCD at small $\mu$ in which the chiral $SO(4) = SU(2)_L \otimes SU(2)_R$ 
symmetry is spontaneously broken down to $SO(3) = SU(2)_{L=R}$ corresponds to 
the antiferromagnetic N\'eel phase of the undoped cuprates in which the 
$SO(3)_s$ spin rotational symmetry is broken down to $SO(2)_s$ due to the 
spontaneous generation of a staggered magnetization. The high-temperature 
superconducting phase of the doped cuprates with spontaneous $U(1)_{em}$ 
breaking corresponds to the color superconducting phase of two flavor QCD in 
which the $SU(3)_c$ gauge symmetry is expected to break down to $SU(2)_c$. 
Finally, the quark-gluon plasma corresponds to the high-temperature metallic 
phase of the cuprates. 

QCD in the color superconducting phase is a genuine high-temperature 
superconductor. The mechanism that leads to quark Cooper pair binding is 
direct one-gluon exchange in the attractive color anti-triplet channel. This is
in contrast to ordinary (low-temperature) superconductors in which the direct 
electron-electron Coulomb interaction (mediated by one-photon exchange) is 
repulsive. Ordinary superconductivity is due to an indirect phonon-mediated 
attraction which occurs at rather low energies and thus gives rise to small 
transition temperatures. The mechanism that leads to high-temperature 
superconductivity in the cuprates is presently not understood, but is expected 
to be due to processes that happen at a rather high energy scale unrelated to 
phonon exchange. Both in the cuprates and in QCD the phase transition that 
separates the phases of broken global and local symmetries is driven by a 
chemical potential (for electrons and quarks, respectively).

Furthermore, the solution of microscopic models for the cuprates (such
as the Hubbard model) is prevented by a severe sign problem. Still,
there are interesting attempts to understand superconductivity in the
cuprates in analogy to simpler condensed matter systems. In
particular, Zhang has conjectured that a transition that separates the
antiferromagnetic N\'eel phase from the high-temperature
superconducting phase is analogous to the spin flop transition of the
staggered magnetization in a 3-d anisotropic antiferromagnet
\cite{Zha90}. This transition is driven by an external uniform magnetic field 
$B$ which plays the role of a chemical potential. Due to the
anisotropy, such an antiferromagnet has only a $\Z(2) \otimes SO(2)_s$
(not the full $SO(3)_s$) spin rotational symmetry. At small $B$, the
staggered magnetization $\vec n = (n_1,n_2,n_3)$ points along the easy
3-axis, while at large $B$ it flops into the 12-plane. The spin flop
transition is illustrated in figure 1.
\begin{figure}[ht]
\begin{center}
\psfig{figure=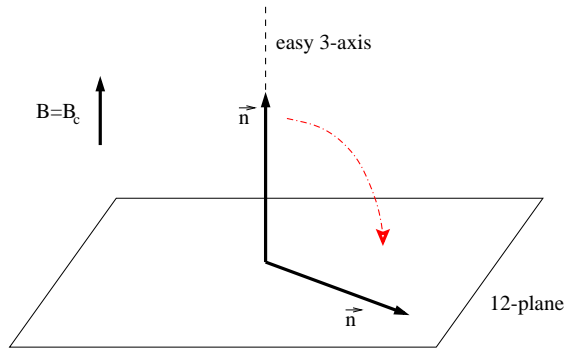,width=7.5cm}
\end{center}
\caption{\it The spin flop transition of an anisotropic antiferromagnet in a
magnetic field $\vec B$. For $B < B_c$ the staggered magnetization vector 
$\vec n$ points along the easy 3-axis, and for $B > B_c$ it flops into the 
12-plane.}
\end{figure}
The flop transition is a first order phase transition line that ends in a 
bicritical point from which two second order phase transition lines emerge --- 
one in the 3-d Ising and one in the 3-d XY model universality class. At the 
bicritical point the $\Z(2) \otimes SO(2)_s$ symmetry is dynamically enhanced 
to $SO(3)_s$ \cite{Fis74}.

Zhang has argued that a similar type of symmetry unification may occur for
high-temperature superconductors \cite{Zha90}. He combined the 3-component 
staggered magnetization and the 2-component Cooper pair condensate to an 
$SO(5)$ ``superspin'' vector $\vec n = (n_1,n_2,n_3,n_4,n_5)$. In the $SO(5)$ 
theory the transition between the antiferromagnetic N\'eel phase and the 
high-temperature superconducting phase is a first order superspin flop 
transition. At small doping (small $\mu$) the superspin lies in the 
$SO(3)_s/SO(2)_s = S^2 $ easy sphere describing the staggered magnetization 
vector. At larger $\mu$ the superspin flops into the $U(1)_{em} = S^1$ plane 
now describing the Cooper pair condensate. The superspin flop transition is 
expected to end in a bicritical point from which two second order lines emerge 
--- one in the 3-d Heisenberg and one in the 3-d XY universality class. Zhang 
has argued that the bicritical point has a dynamically enhanced $SO(5)$ 
symmetry although the microscopic Hamiltonian is only 
$SO(3)_s \otimes U(1)_{em}$ invariant. The expected phase diagram of an 
$SO(N) \otimes SO(M)$ invariant theory with potential dynamical symmetry 
enhancement to $SO(N+M)$ is shown in figure 2. For the cuprates $N = 2$ and 
$M = 3$, while for anisotropic antiferromagnets $N = 2$ and $M = 1$. As we will
see next, for QCD $N = 6$ and $M = 4$.
\begin{figure}[ht]
\begin{center}
\psfig{figure=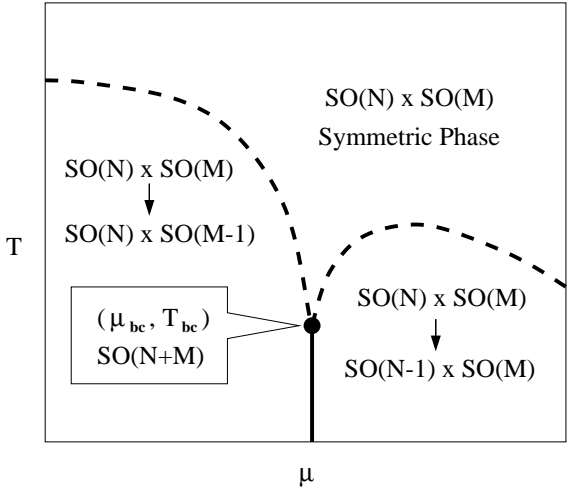,width=7.5cm}
\end{center}
\caption{\it Expected phase diagram of an $SO(N) \otimes SO(M)$ invariant 
theory with potential dynamical symmetry enhancement to $SO(N+M)$. The first 
order flop transition (solid) line ends in a bicritical point 
$(T_{bc},\mu_{bc})$. Two second order (dashed) lines emerge vertically from 
this point.}
\end{figure}

Recently, we have generalized the $SO(5)$ unified theory of high-temperature 
superconductivity and antiferromagnetism to an $SO(10)$ unified description of 
color superconductivity and chiral symmetry breaking in QCD \cite{Cha00}. 
Although the unifying group is the same as in a grand unified theory, the 
unification scale would now be around 10 MeV. We consider left and right-handed
quark fields $\Psi_L^{f,c}$ and $\Psi_R^{f,c}$ with two flavors $f = 1,2$ and 
three colors $c = 1,2,3$. The chiral symmetry breaking order 
parameter 
\begin{equation}
(\Psibar\Psi)^{fg} = \sum_c \Psibar_L^{f,c} \Psi_R^{g,c}
\end{equation}
is a color singlet, $SU(2)_L$ and $SU(2)_R$ doublet, with baryon number zero. 
The color symmetry breaking order parameter
\begin{equation}
(\Psi\Psi)^c = \sum_{f,g,a,b} \epsilon_{fg} \epsilon_{abc} (\Psi_{L,R}^{f,a})^T
C \Psi_{L,R}^{g,b},
\end{equation}
on the other hand, is a color anti-triplet, $SU(2)_L \otimes SU(2)_R$ singlet, 
with baryon number $2/3$. Similarly, $(\Psibar\Psibar)^{c}$ is a color triplet,
$SU(2)_L \otimes SU(2)_R$ singlet, with baryon number $- 2/3$. The group 
$SO(10)$ contains $SU(3)_c \otimes SU(2)_L \otimes SU(2)_R \otimes U(1)_B$ as a
subgroup. The 10-dimensional vector representation of $SO(10)$ decomposes into
\begin{equation}
\{10\} = \{1,2,2\}_0 \oplus \{\bar 3,1,1\}_{2/3} \oplus \{3,1,1\}_{-2/3},
\end{equation}
and thus naturally combines the order parameters for chiral symmetry breaking 
and color superconductivity to a 10-component ``supervector'' $\vec n = 
(n^1,n^2,...,n^{10})$ with 
\begin{eqnarray}
&&n^c =(\Psi\Psi)^c + (\Psibar\Psibar)^c, \nonumber \\
&&n^{c+3} = - i [(\Psi\Psi)^c - (\Psibar\Psibar)^c], \ c \in \{1,2,3\}, 
\nonumber \\
&&n^7 = (\Psibar\Psi)^{11} + (\Psibar\Psi)^{22}, \nonumber \\
&&n^8 = - i [(\Psibar\Psi)^{12} + (\Psibar\Psi)^{21}], \nonumber \\ 
&&n^9 = (\Psibar\Psi)^{12} - (\Psibar\Psi)^{21}, \nonumber \\ 
&&n^{10} = - i [(\Psibar\Psi)^{11} - (\Psibar\Psi)^{22}].
\end{eqnarray}
In the chirally broken phase at small $\mu$ the 4-component vector 
$(n^7,n^8,n^9,n^{10})$ develops an expectation value, thus breaking 
$SU(2)_L \otimes SU(2)_R$ spontaneously to $SU(2)_{L=R}$. The corresponding 
Goldstone pions are described by fields in the 
$SU(2)_L \otimes SU(2)_R/SU(2)_{L=R} = S^3$ easy 3-sphere. In the color 
superconducting phase at larger $\mu$, on the other hand, the 6-component 
vector $(n^1,n^2,...,n^6)$ gets an expectation value and the supervector flops 
into the 5-sphere $SU(3)_c/SU(2)_c = S^5$ that describes the quark Cooper pair
condensate. In this case, one would expect that the first order supervector 
flop line ends at a bicritical point with dynamical symmetry enhancement to 
$SO(10)$. However, a renormalization group flow analysis shows that both the
$SO(5)$ and $SO(10)$ fixed point are unstable, at least in $(4 - \epsilon)$ 
dimensions \cite{Cha00}. We have simulated the flop transition in simple 
classical $SO(N) \otimes SO(M)$ invariant models. Figure 3 shows the flop of 
the staggered magnetization from the 3-axis into the 12-plane for an 
anisotropic antiferromagnet. Similar studies for high-temperature and color 
superconductors are in progress. In particular, we investigate the endpoint of 
the flop transition in order to see if the perturbative results obtained in 
$(4 - \epsilon)$ dimensions also apply to the nonperturbative 3-d case. 
\begin{figure}[ht]
\begin{center}
\psfig{figure=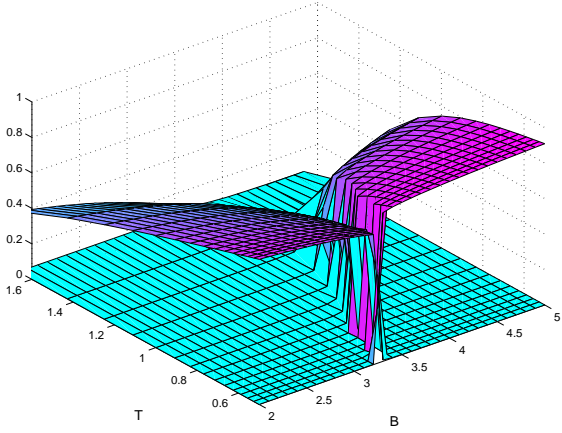,width=7.5cm}
\end{center}
\caption{\it Flop transition in an anisotropic antiferromagnet. The two wings
correspond to the square of the staggered magnetization $n_3^2$ along the 3-axis and 
$n_1^2 + n_2^2$ in the 12-plane.}
\end{figure}

To illustrate the dynamics of the flop transition, we now construct a unified 
low-energy effective Lagrangian for the Goldstone modes described by an
$(N + M)$-component unit vector $\vec n$. In the absence of $SO(N+M)$ symmetry
breaking terms (other than the chemical potential), the low-energy effective 
action takes the form
\begin{eqnarray}
S[\vec n]&=&\int_0^{1/T} dt \int d^3x \ \frac{F^2}{2} 
[\partial_i n^\alpha \partial_i n^\alpha \nonumber \\
&+&\frac{1}{c^2} (\partial_0 n^\alpha + A_0^{\alpha \beta} n^\beta)
(\partial_0 n^\alpha + A_0^{\alpha \beta} n^\beta)].
\end{eqnarray}
The chemical potential $\mu$ couples as an imaginary non-Abelian constant 
vector potential 
\begin{equation}
A_0^{\alpha\beta} = i \mu \!\! \sum_{c = 1,...,N/2} \!\!
(\delta^{\alpha,c} \delta^{c+N/2,\beta} - 
\delta^{\alpha,c+N/2} \delta^{c,\beta})
\end{equation} 
in the Euclidean time direction. To account for explicit $SO(N+M)$ breaking to 
$SO(N) \otimes SO(M)$ we add a potential term 
$- V_0 [(n^{N+1})^2 + ... + (n^{N+M})^2]$ to the action that favors the easy 
$(M-1)$-sphere. In the case of QCD this leads to chiral symmetry breaking. The 
total potential for constant fields $\vec n$ then takes the form
\begin{eqnarray}
V(\vec n)&=&- \frac{F^2}{2 c^2} \mu^2 [(n^1)^2 + ... + (n^N)^2] \nonumber \\
&-&V_0 [(n^{N+1})^2 + ... + (n^{N+M})^2].
\end{eqnarray}
For $\mu < \mu_c = \sqrt{2 V_0 c^2/F^2}$, it is energetically favorable for
the supervector to lie in the easy $(M-1)$-sphere. For $\mu > \mu_c$, on the 
other hand, the supervector $\vec n$ flops into the $(N-1)$-sphere. In QCD, this
induces a first order phase transition from the chirally broken to the color 
superconducting phase.

It is interesting to ask if the supervector can play a dynamical role in the 
real world. In particular, with the strange quark present, a new color 
superconducting phase with color-flavor locking arises \cite{Alf99}. This phase
may be analytically connected to the ordinary hadronic phase \cite{Sch99}. Then
there would be no supervector flop transition. However, when the strange quark 
is sufficiently heavy, a flop transition may exist.

\section*{Acknowledgements}

This work is supported in part by funds provided by the U.S. Department of 
Energy (D.O.E.) under cooperative research agreements DE-FC02-94ER40818 and 
DE-FG02-96ER40945. U.-J. W. also acknowledges the support of the A. P. Sloan 
foundation.

\end{document}